\documentstyle[12pt,aaspp4,psfig]{article}
\def\ltsima{$\; \buildrel < \over \sim \;$}
\def\simlt{\lower.5ex\hbox{\ltsima}}
\def\gtsima{$\; \buildrel > \over \sim \;$}
\def\simgt{\lower.5ex\hbox{\gtsima}}
\def\kms{{\rm\,km\,s^{-1}}}
\def\kpc{{\rm\,kpc}}

\def\yr{{\rm\,yr}}

\def\s{\ifmmode \widetilde \else \~\fi}
\def\={\overline}

\def\spose#1{\hbox to 0pt{#1\hss}}

\def\etal{{\it et al.\ }}

\def\eg{{ e.g.,\ }}

\def\lta{\mathrel{\spose{\lower 3pt\hbox{$\mathchar"218$}}
     \raise 2.0pt\hbox{$\mathchar"13C$}}}
\def\gta{\mathrel{\spose{\lower 3pt\hbox{$\mathchar"218$}}
     \raise 2.0pt\hbox{$\mathchar"13E$}}}
\def\Dt{\spose{\raise 1.5ex\hbox{\hskip3pt$\mathchar"201$}}}	
\def\dt{\spose{\raise 1.0ex\hbox{\hskip2pt$\mathchar"201$}}}	

\def\=={\equiv}

\def\dotsfill{\leaders\hbox to 1em{\hss.\hss}\hfill}

\slugcomment{Accepted by The Astronomical Journal}

\lefthead{Ibata \& Lewis}
\righthead{Optimal proper motions with WFPC2}

\begin{document}

\title{Optimal proper motion measurements with  the Wide Field and Planetary
Camera}

\author{
Rodrigo A. Ibata\altaffilmark{1}, 
Geraint F. Lewis\altaffilmark{2}}

\altaffiltext{1}
{European Southern Observatory 
Karl Schwarzschild Stra\ss e 2, D-85748 Garching bei M\"unchen, Germany \nl
Electronic mail: ribata@eso.org}

\altaffiltext{2}{ 
Fellow of the Pacific Institute for Mathematical Sciences 1998-1999, \nl
Dept. of Physics and Astronomy, University of Victoria, Victoria, B.C., Canada 
\& \nl 
Astronomy Dept., University of Washington, Seattle WA, U.S.A.
\nl
Electronic mail: {\tt gfl@uvastro.phys.uvic.ca} \nl 
Electronic mail: {\tt gfl@astro.washington.edu}}



\begin{abstract}
An optimal  maximum-likelihood  technique for computing  point-source  image
centroids from  many, slightly offset,  CCD frames is presented.  The method
is especially  useful for measuring  stellar proper motions  from data taken
with the Wide Field and Planetary Camera aboard the HST, and also provides a
means   to identify very  compact non-stellar  sources.  We  work though the
example problem of  obtaining image centroids of objects  in the Hubble Deep
Field.
\end{abstract}




%

\section{Introduction}

The Wide    Field and  Planetary  Camera  (WFPC2)  aboard  the  Hubble Space
Telescope (HST),  is  an excellent  tool for astrometric  study.   This is a
result of  its superb spatial  resolution and image stability.  However, the
Wide Field  (WF) camera produces  undersampled images, and  the point spread
function (PSF) of  WFPC2 is complex. These  properties force one to  have to
take special care in the data reduction process.

In the  present    contribution we  discuss   an  optimal maximum-likelihood
technique that  is   particularly well-suited for  calculating  centroids of
point-like objects   from   WFPC2 data  obtained  in  many  slightly  offset
exposures.  The same technique  will be readily  applicable to STIS  imaging
data or Advanced Camera images.

As a working  example, we present  the problem of obtaining image  centroids
for point-like objects in the Hubble Deep Field (HDF).

\section{Astrometry with WFPC2}

The expected proper motion (PM) $\alpha$ for a point-source at distance $d$,
moving with  transverse velocity of $v_{\perp}$ over  a time interval of $T$
is  
$\alpha  =  1.79   {{v_{\perp}} \over {200\kms}}  
{T \over {2\yr}} 
({d \over {1\kpc}})^{-1}
{\rm pixels}$ 
on the Planetary Camera (PC) and 
$\alpha  =  0.82   {{v_{\perp}} \over {200\kms}}  
{T \over {2\yr}} 
({d \over {1\kpc}})^{-1}
{\rm pixels}$ 
on the WF   camera  (the velocity, distance   and time  interval values  are
appropriate  to our  test  problem below).  So the  HST  guiding accuracy of
$0.005$  arcsec RMS (corresponding to $0.1$  PC pixels and $0.05$ WF pixels)
in "fine lock" mode, is sufficient to measure  the PMs of Galactic stars out
to  large  distances, as  long  as accurate image   centroids and a suitable
reference  frame  can be  determined to better    accuracy than the expected
proper motions.   This jitter of  0.005 arcsec,  will make  the undersampled
WFPC2 images fractionally wider, but it will  not significantly affect image
centroiding (to  better than the RMS  uncertainty), unless  the jitter has a
systematic direction.

The WFPC2 dithering  technique, where  many exposures of  a  field are taken
with slightly different pointings, offset by a  few pixels, was developed as
a  means to eliminate the effects  of CCD cosmetic  defects,  hot pixels and
cosmic rays.  It is  also very useful for   astrometric purposes, since  the
positional accuracy of each  dithered frame allows  the stellar image, which
is undersampled in  a  single frame, to  be  resampled at  several sub-pixel
positions.  Specialized software,  such   as the ``drizzle''    algorithm of
\markcite{fru97}Fruchter \& Hook (1997), have been devised to stack dithered
frames.     Taking advantage  of    the extra   positional information, they
dramatically enhance the resolution of the final stacked image.

However, stacking inevitably degrades information.   The resulting PSF  must
always be more complex  than that of individual  frames, and given that  the
WPFC2 PSF varies strongly as a function of position over the camera, the PSF
of  the  stacked image will  also  depend on the  particular  dither pattern
adopted by the  observer. These problems become  more severe if one enhances
the resolution of the  stacked image (with  such algorithms as ``drizzle'');
furthermore,   the noise   in the   stacked image   will then  be  spatially
correlated and hence quite complex.  On  the other hand, working purely with
shallow individual frames is  a huge waste of the  depth of the data-set (in
the HDF,  objects that have  S/N$\sim 8$ in  the  combined 58-exposure first
epoch F814W stack, correspond   to $\sim 1\sigma$ detections  in  individual
exposures).

The solution to this apparent dilemma is very simple.  We  assume the PSF in
each frame  (and its variation across each  frame) is known.  These PSFs are
best determined from the data-set under investigation.
\footnote{If there is a systematic  direction to the telescope jitter  which
changes from  exposure to exposure, its effect  can be largely eliminated by
measuring the PSF of each data-frame individually.}
However, if there are few or no  bright, isolated, unsaturated stars present
in those frames, one may still be able to obtain a good approximation to the
PSF, if  it sufficiently stable  over time. This is  the case for WFPC2; and
suitable data  are readily obtained from the  archive.  We  also assume that
the  transformations (and inverse transformations)  that map  every point on
the  $i$th frame  to a Cartesian   grid on the sky  have  been determined in
advance   (using  techniques such as  those   described  below).  A ``master
frame'',  whose rows and  columns are  aligned with  that Cartesian grid  is
produced   by stacking all  the   individual frames  (using  the ``drizzle''
algorithm, for instance, but  with resolution enhancement  turned  off).  On
this ``master frame'', we search for  candidate stars and photometer them; a
crowded-field       photometry  package       such      as       ``ALLSTAR''
(\markcite{ste94}Stetson 1994) is ideal for this  purpose.  This also yields
a first  estimate $(x_j,y_j)$ of   the astrometric position on the  ``master
frame''  of  the   $j$th  candidate   point-source.   These  positions   are
transformed into the   coordinate  system   of  the  $i$th  frame to    give
$(x^i_j,y^i_j)$.  Then, for each of the $j=1,  \dots, N$ stellar candidates,
we find the likelihood of the the exposure-normalized PSF model in the $i$th
frame, given the data $D^i$ in the $i$th frame
$$L^i_j(x^i_j,y^i_j) = \prod_{k,l} L[{\rm PSF}^i(k,l)  + S^i_j | D^i(k,l) ],
\eqno(1)$$
where the product is performed over all uncontaminated pixels $k,l$ within a
circle of radius $R$ of  $(x^i_j,y^i_j)$.  Contaminated pixels --- by  which
we mean pixels on cosmetic defects, hot pixels, or pixels affected by cosmic
ray impacts --- are simply left out of the calculation (we discuss below how
bad pixel  maps were constructed for each  frame of our  test data-set). The
quantity $S^i_j$ is  the modal sky value in  an annulus with suitably chosen
inner  and outer radii  around the $j$th object on   frame $i$.  The product
over all $M$ frames of $L^i_j$,
$$L_j(x_j,y_j) = \prod_{i=1, \dots, M} L^i(x^i_j,y^i_j),$$
is the likelihood that a point-source is centered at position $(x_j,y_j)$ on
the ``master frame'', given all the available  data.  Repeating this process
in   a  fine  grid  of  $(x,y)$  values  in  the immediate   neighborhood of
$(x_j,y_j)$, yields   the   likelihood   surface, and   a    two-dimensional
maximization routine can  then be used to find  the  position $(X_j,Y_j)$ of
the maximum of  $L_j$. The coordinate   $(X_j,Y_j)$ is then the  most likely
position of  the center of the image.   There has been no image degradation,
as the data have not  been tampered with  (except for initial debiassing and
flat-fielding), and  there is no loss  of depth.  The technique  is optimal,
and  is  especially useful for the   case where rotated, optically distorted
frames with different and (or) spatially varying PSFs are to be analyzed.

The important noise sources are: Poisson noise  in the source, Poisson noise
in the  sky, read noise,   flat-fielding errors and PSF mismatching  errors.
When dealing  with  faint objects  on  a  low  sky  background  it  will  be
advantageous to consider  carefully   the distribution of  expected  counts,
which is why we stressed the use of the likelihood function above.  However,
for all the images we analyzed, the sky  background was substantial, greater
than 100~${\rm e^-}$, so  that  the noise  distribution could  be reasonably
modeled by a Gaussian error distribution on each pixel.  Given this, one may
then trivially  compute,  using   the  $\chi^2$ statistic  instead  of   the
likelihood in  the computations  above,  the  probability that  the observed
brightness enhancement is drawn from the same distribution as a point-source
located  at $(X_j,Y_j)$. Thus, this  method also provides an excellent means
of discriminating point-like  from   extended sources, that   is  especially
powerful at revealing objects that deviate only slightly from point-sources.
Again, this   uses the full  depth  of the data-set,  avoids the information
degradation inherent to  the stacking process,  and also avoids the problems
of having correlated noise.

We found that a considerable improvement in $\chi^2$ can  be achieved if the
magnitude as well as   the position of   bright candidate point-sources  are
refined simultaneously.    Though the flux estimate did   not vary  from the
input value by more than  0.1 magnitudes for any of  the objects we measured
in the test problem below, the $\chi^2$ probability occasionally improved by
orders of magnitude.

Although we have not tested  this technique on crowded  fields, it is likely
to yield  accurate centroid positions if the  input positions and magnitudes
of   all detectable stars   have  been  carefully determined   with  a  good
crowded-field photometry package. However, one should make the following two
alterations to the algorithm.  First, before analyzing the $j$th object, one
should subtract the  expected counts from all other  objects from each frame
in the data set.   And second, the  objects should preferably be analyzed in
order of decreasing flux.

A limitation of our method is that we choose not to measure variability over
the time-span of the observations in any one epoch. Clearly, adding an extra
parameter for  each  object  on each frame,    would make  the  scheme  less
robust. Variability between epochs can be measured, however.

\section{An example: Proper motions in the Hubble Deep Field}

The HDF  is the deepest image yet  obtained of the  Universe; taken with the
Hubble Space  Telescope (HST) in  director's discretionary time  in December
1995 (\markcite{wil96}Williams \etal\ 1996), its primary aim of studying the
formation an evolution of galaxies has been  extremely successful.

Apart from the numerous galaxies, a small number of stars were also detected
in this field.  However,  the constraints that can  be placed  upon Galactic
structure models from  these data are  disappointing  due to  the relatively
shallow limit of ${\rm I=26}$  at which stars  can be discriminated from the
galaxies  with reasonable  confidence  (\markcite{fly96}Flynn \etal\  1996).
For comparison, the limiting magnitude of  the HDF in  F814W is ${\rm I \sim
28}$, so a factor of $\sim 15$ of survey volume would be gained if one could
push star-galaxy discrimination to the faintest limit of the data.

Here we show that a better way to find stars at the faint limit of this data
set is to observe the field in F814W in a second epoch, and calculate proper
motions (PMs).  The  necessary follow-up data, taken  in December 1997, were
obtained   to   undertake   a     search  for    high  redshift   supernovae
(\markcite{gil98}Gilliland \&  Phillips 1998).  Note  that over the two year
timespan that separates the datasets, a star $1 \kpc$ away travelling with a
transverse velocity  of $200  \kms$ will  move 0.82  WF  pixels.  Transverse
velocities   of this   magnitude  can  be expected  from  spheroid  (or more
interestingly) halo stars due  to  the large  velocity dispersions  of these
populations.  Furthermore, their   slow rotation about  the Galactic  center
will introduce a large apparent motion as viewed from  the Sun.

\subsection{Registration of frames}

To apply  the   method outlined above,   we  need  to know the   geometrical
transformations between pixel  positions  on each  frame  taken at a   given
epoch.   To  do this,  we  could find the  centroids  of  objects  (stars or
galaxies) on each frame  and compute transformation  coefficients.  However,
due to optical  design, WFPC2 images  give a substantially distorted view of
the sky,   so  a high  order  polynomial must   be  used to  give acceptable
residuals.   If  there are few   objects on the   frames  for which reliable
centroids can be found, this procedure will be far from robust. It is better
therefore to make use of  some prior information:  the optical distortion of
WFPC2,   as  a  function  of  wavelength,    is   fairly  well   understood.
\markcite{tra95}Trauger \etal\  (1995)  have   published transformations  (a
10-coefficient bicubic polynomial  in each of the  $x$ and  $y$ directions)
that  allow one to convert CCD  pixel positions, in  a given  passband, to a
geometrically corrected frame.  The accuracy of this transformation has been
determined to be 0.1 pixels RMS over the fields of view of the CCDs.

Though not necessary, the proper motion analysis is easier if one is able to
construct an accurate extragalactic reference frame. Ideally, one would like
to use  point-like sources for this purpose,  as the centroid of an extended
source (which is most probably also lumpy) is  not easily defined.  However,
to date no QSOs have been identified in the HDF.  One is therefore forced to
use galaxies to define the reference frame. 

Galaxies are generally fuzzy, lumpy objects, so it is very difficult, if not
perhaps  impossible, to define  the  center of  the light  distribution.  To
circumvent  this   problem,  the approach    we  take   is  to  obtain  {\it
differential}  measurements.  Only   for  the  purpose  of   determining the
transformations between frames,   we ``un-distort'' all   of the  frames  by
applying the \markcite{tra95}Trauger \etal\ (1995) coefficients.  Bad pixels
in  the  frames are flagged, as  described  below.  The dither  position \#6
frames  (see \markcite{wil95}Williams \etal\  1996)  were chosen to define a
``reference frame''.  A  first estimate of the   positions of the  $\sim 50$
brightest  galaxies in the chosen  ``reference frame'' is obtained using the
``FIND'' algorithm of ``DAOPHOT'' (\markcite{ste87}Stetson 1987), which fits
a  Gaussian function to brightness  enhancements in the image.  However, the
resulting positions are accurate to not much better than  about 1 pixel.  To
improve this positional accuracy, we implemented the following algorithm.

First, a search is  performed for the local minima  of the  marginal density
distributions along the column and row directions of a 40 pixel box centered
on each input position.  This is done in  two iterations following the first
steps of  the recipe given  in \markcite{ste79}Stetson (1979).   These local
minima are found on  either side of  the peak, both  in  in the $x$ and  $y$
directions, and are used to define the  limits of a  new box surrounding the
object  under study.  This box will   be free of   the influence of brighter
neighbors.  The marginal density distributions in this new box are computed.
The  algorithm then cross-correlates  the  marginal density distributions of
the same object on each frame.  Obviously, best  results will be obtained if
the frames being compared have approximately the same orientation
angle.
\footnote{We  also experimented performing two-dimensional cross-correlation
of the cut-out images.   The accuracy achieved  is approximately the same as
in  the  procedure outlined above,  but less  robust, in   that wildly wrong
answers were  often returned when  the images did   not have high  signal to
noise.}

Having determined the positional offsets for  all the bright galaxies on all
frames,  one can proceed to find  the geometric  transformations that relate
the frames to each other.  A simple four-coefficient geometric transformation
(shift, rotation and change of scale) was found to give excellent residuals,
better than $0.02$ pixels RMS, between frames in the same passband. Thus the
Trauger   \etal\  model  provides an  accurate   map  of the  WFPC2  optical
distortion. (The fact that we  are able to  align the undistorted frames  to
better than the   expected accuracy  of  the distortion  transformations  is
probably a consequence  of the relatively close  alignment of the individual
HDF frames).

This procedure has given   us   the geometric transformations between    the
(undistorted) ``reference frame'' and all other undistorted frames. However,
what  we really  need     to know  are  the   transformations   (and inverse
transformations)  between  the  raw  (optically  distorted) frames  and  the
``reference frame''.  The   forward transformations   are found  by   simply
substituting the  Trauger  \etal\ functions into  the above four-coefficient
geometric transformations.  The inverse   transformations cannot be  written
down as a polynomial, but a  non-linear Newton-Raphson algorithm can be used
to provide the required inverse mapping.

\subsection{Bad pixel rejection}

Finally,  it is necessary  to flag bad pixels   on the raw frames.  Consider
pixel  $k,l$ on   the $i$th  frame.  We  find  the overlap  area  (using the
computed geometrical transformations) of the  footprint of this pixel on all
other frames in  the same passband as  frame $i$ and in  the  same epoch, to
obtain a list of $M$ flux estimates (i.e.  counts  per second above the sky)
at this  position.  The   mean and  standard  deviation  of  this list   are
computed, after clipping the highest flux datum.  Given the exposure lengths
and the number of frames, this  datum is likely  to be severely contaminated
by a  cosmic  ray; however,  the statistical  bias introduced   by this clip
should be  negligible.  The  pixel is flagged  as  bad if the  measured flux
deviates from the  mean flux by  more  than four  standard deviations.  This
process is  repeated for all  pixels on all frames.   The  advantage of this
procedure is clear: maximum spatial resolution is maintained on all frames.

\subsection{Sample results}

As an illustration, in Figure~1  we show the  result of applying the present
technique to three sample objects on the WF2 chip  using data from the first
epoch HDF.   The upper two panels  show results relating  to  a fairly faint
star identified by   \markcite{fly96}Flynn \etal\ (1996)  (${\rm  I=24.78}$,
${\rm V-I=1.64}$; position on HDF WF2 dither  \#6: $x=350$, $y=469$).  Panel
(a) shows  the likelihood contours  of  the centroid  of  this star (in  the
``reference frame''), calculated using all first epoch F814W exposures.  The
``star'' graph-marker shows the point  $(X_j,Y_j)$, the most likely centroid
position, while the $n$th contour marks the boundary of the region where the
likelihood has fallen by a factor of $\exp{-{{n^2} \over  2}}$ from the most
likely value (so the image centroid is $< 10^{-22}$  times less likely to be
situated  beyond  the last  contour  than at the  position   of the ``star''
marker).   The distance from  the  ``star'' marker  to  the first contour is
0.017 pixels,  or 1.7~mas.   Panel (b)  displays  the  object profile.   The
position  $(X_j,Y_j)$  is transformed   to   the correct  position  on   the
individual raw frames, $(X^i_j,Y^i_j)$; for all frames $i$,  we plot the raw
pixel data within 2 pixels of $(X^i_j,Y^i_j)$ as a function of distance from
that  point.  The   uncertainties   on individual   pixel values   are  also
indicated.  The expected  counts  from the  PSF models are  shown as  filled
circles; that  these values  do  not always  decrease monotonically from the
image center  is  due  to non-axisymmetry in   the  model PSFs,  and  to the
particular way  in  which the dithering sampled   the object.   This diagram
serves simply to illustrate the goodness of  fit of the data  to the PSF; we
find that the probability that $\chi^2$, for a correct model, should be less
than the observed value is $P=0.47$.

The middle  two panels display the  results of applying  the technique  to a
faint, blue star  identified  by \markcite{fly96}Flynn \etal\  (1996) (${\rm
I=26.22}$, ${\rm V-I=0.16}$;    position on HDF  WF2  dither  \#6:  $x=322$,
$y=637$).  Panels (b) and (c)  have, respectively, similar content to panels
(a) and (b). Here,  the centering uncertainty  has degraded to $\sim 3$~mas.
Using the $\chi^2$ statistic, we find $P=1.9\times  10^{-4}$, so this object
is almost certainly not a point-source,  illustrating the resolving power of
our  method.  Interestingly,   other   faint  blue  objects   identified  by
\markcite{fly96}Flynn  \etal\ (1996) as stars  can similarly be  shown to be
non-stellar;   the  nature  of these  objects    will  be investigated in  a
subsequent contribution.

Finally, panels (d) and (e) show the  results for an  object at the limit of
the HDF,  (${\rm I=27.9}$, no color information  available;  position on HDF
WF2  dither \#6: $x=326$, $y=250$).  Our  estimated  centroiding accuracy on
this extremely faint object is $\sim 10$~mas.

\subsection{Positional accuracy}

Many  sources contribute to uncertainty in  the computed centroid positions.
There are ``fundamental'' uncertainties from Poisson noise in the sky and in
the object, from detector noise, and from the  sampling.  There will also be
uncertainties  in the flat-fielding,  and in the  PSF determination. Further
sources of uncertainty, not accounted for in our model,  arise from the fact
that CCD  pixels are not  exactly square, that  they are  not laid out  on a
perfect   Cartesian grid,  and  that,  at  some  level,  every  pixel  has a
non-uniform sensitivity over its surface.

Clearly, it is  desirable  to determine the  combined  effect of all   these
uncertainties.   To this end,  we  separated the  first epoch  F814W data at
dither positions 1-5  and  at dither  positions 6-11 \footnote{data  was not
obtained    at   all  dither   positions   in  all   filters}    to make two
sub-samples. Comparing  the positions of   point-sources determined from the
first  sub-sample  ($\vec{x_1}$,  say) to those    determined  in the  other
($\vec{x_2}$), provides an  internal  means to measure  the accuracy  of the
method.  The HDF frames  are slightly shifted  and some are slightly rotated
with respect to each other, so this exercise should provide a good indication
of the achievable centering (and proper motion) accuracy for a dataset where
the frames of all epochs are in close alignment.

This experiment  was performed   on   the seventy-two objects, with    light
profiles consistent with being point-sources with probability $P>0.01$, that
we detected in the HDF. The results are displayed in Figure~2.  The ``star''
graph-markers show the value of ${{|\vec{x_1}-\vec{x_2}|} \over {\sqrt{2}}}$
as a function of ${\rm I}$~magnitude.  The filled  circles give the expected
uncertainties,  as derived from the  likelihood surfaces. The good agreement
between   these  two methods  of   estimating the  centroiding uncertainties
suggests that  our noise  model is reasonable.   Clearly,  it is possible to
obtain quite accurate  centroids (to $\sim 10$~mas) even  at the very  faint
limit of the HDF, opening the possibility of many interesting studies.

\section{Conclusions}

A  method  has been  outlined    for obtaining accurate  point-source  image
centroids    from    WFPC2  data.   It  is   optimal,   in   the sense  that
maximum-likelihood techniques are  used  to take advantage of  all available
positional and brightness data contained in the CCD frames.

It is shown that,  when applied to  the HDF data-set, centroid uncertainties
on the order of 2~mas are easily  achieved for relatively faint stars, while
stars at the limit of the data-set, near ${\rm I  \sim 28}$, may be measured
with accuracies of  $\sim 10$~mas.  Many  interesting proper motion  studies
are  therefore  possible.

Further  work is required  to determine up to what  accuracy the bulk proper
motion of point-sources (\eg\ star cluster or local group galaxies) improves
as the square root of the number of sources.   Systematic effects must drown
the signal at some level; but judging from  the present work, our 0.02 pixel
limit is set by the  accuracy with which  we were able  to fix the reference
frame.  The bulk proper motions of even very faint populations can therefore
be measured down to at least that  level of accuracy.   So for projects that
require   proper motion measurements   of  very faint  sources, HST  imaging
instruments are likely  to remain highly competitive   even compared to  the
next generation of astrometric  missions such as NASA's Space Interferometry
Mission  (SIM) or ESA's   Global Astrometric Interferometer for Astrophysics
(GAIA).

The  proper motions measured  between the original HDF  and the second epoch
exposures in that  field will  be analyzed and  presented in  a  forthcoming
paper (\markcite{me98}Ibata \etal\ 1998).

\clearpage

\begin{figure}
\psfig{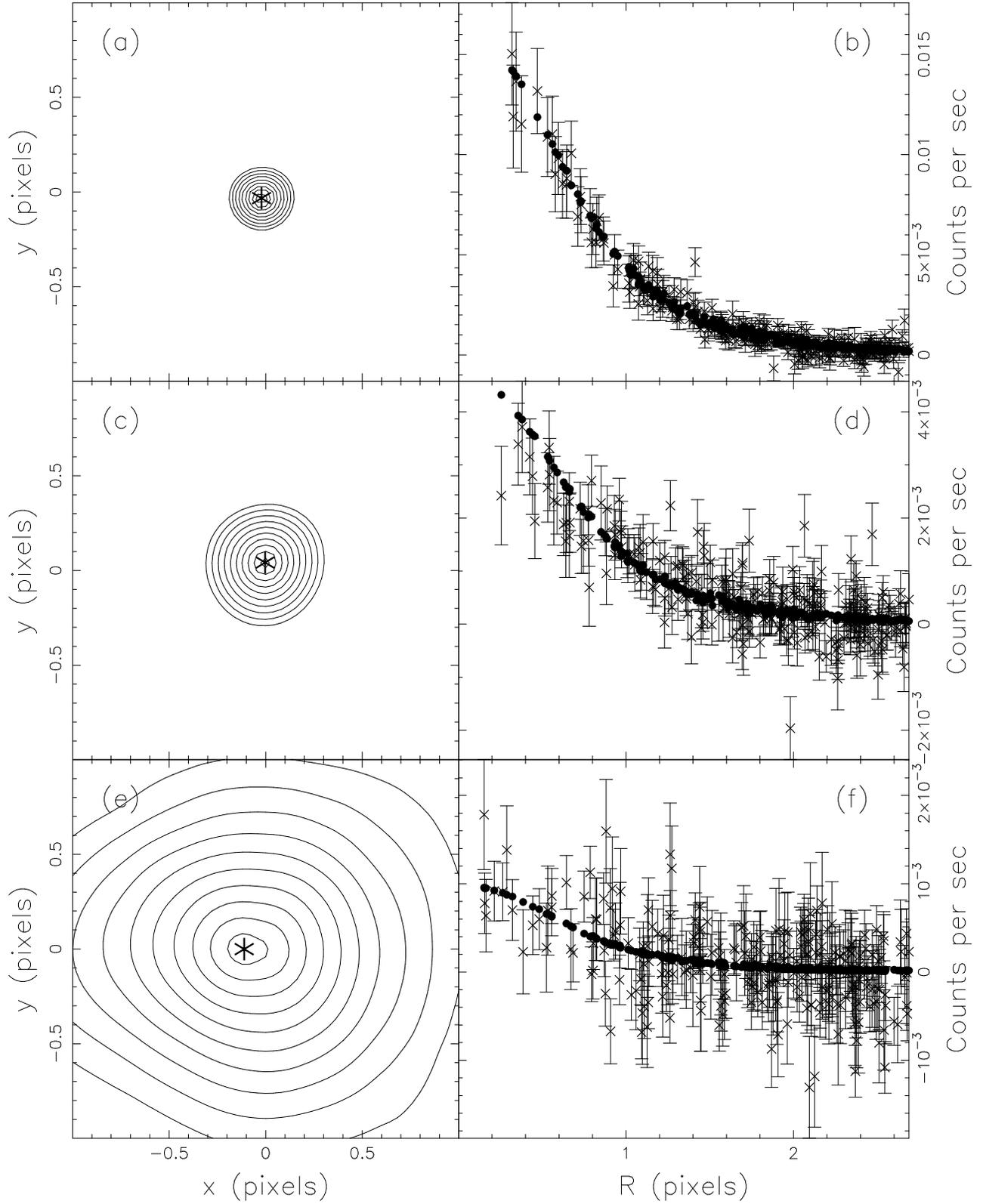}    
\figcaption[Ibata.fig01.ps]{The centroid  likelihood surfaces   and    image
profiles of  three stars  of   magnitude I=24.78,  I=26.22,  and I=27.9  are
displayed, respectively,  in the upper, middle and  lower panels. A detailed
explanation of these diagrams is given in the text.}
\end{figure}
\clearpage

\begin{figure}
\psfig{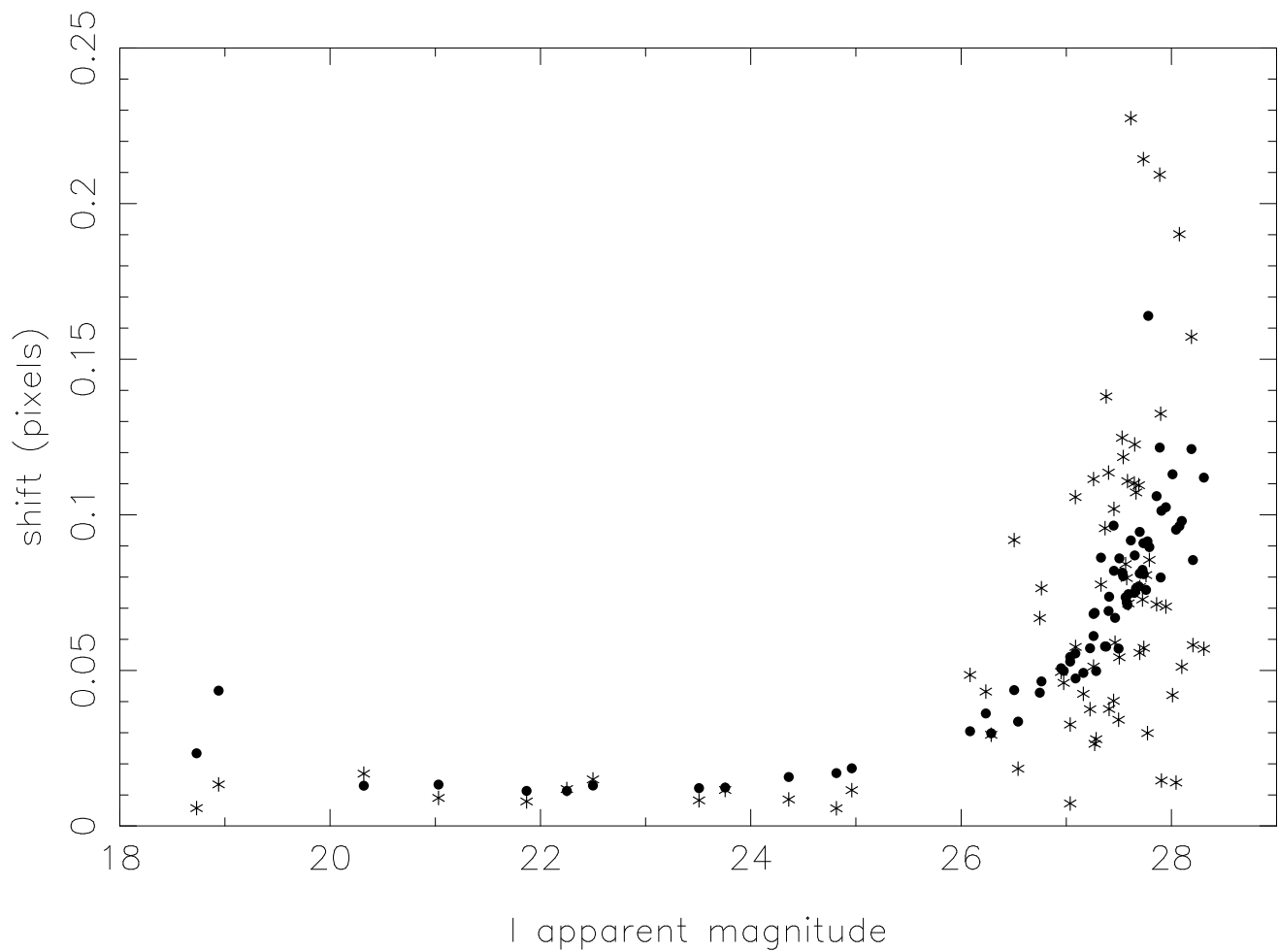} 
\figcaption[Ibata.fig02.ps]{The  uncertainty in   the centroid  position  of
point-like objects on the HDF WF chips is displayed  as a function of I-band
apparent   magnitude.   The ``star''  markers show ${{|\vec{x_1}-\vec{x_2}|}
\over {\sqrt{2}}}$, that is, ${1/\sqrt{2}}$    times the difference in   the
centroid position derived from two, approximately equal exposure, subsamples
of   the original    HDF  dataset.  The filled-circles    show  the expected
uncertainties, derived from the centroid likelihood surfaces.}
\end{figure}

\end{document}